# pH-jumps in a Protic Ionic Liquid Proceed by Vehicular Proton Transport

Sourav Maiti[§], Sunayana Mitra[¶], Clinton A. Johnson[¶,$], Kai C. Gronborg[¶], Sean Garrett-Roe[¶*] and Paul M. Donaldson[§*]

[§]*Central Laser Facility, RCaH, STFC-Rutherford Appleton Laboratory, Harwell Science and Innovation Campus, Didcot, United Kingdom*

[¶]*Department of Chemistry, University of Pittsburgh, 219 Parkman Avenue, Pittsburgh, Pennsylvania 15260, USA*

*Corresponding Authors

Sean Garrett-Roe: sgr@pitt.edu

Paul M. Donaldson: paul.donaldson@stfc.ac.uk

[$]Current address: *Department of Chemistry, Davis and Elkins College, 100 Campus Dr., Elkins, WV 26241, USA*

**Abstract**

The dynamics of excess protons in the protic ionic liquid ethylammonium formate (EAF) have been investigated from femtosecond to microseconds using visible pump mid-infrared probe spectroscopy. The pH-jump following visible photoexcitation of a photoacid (8-hydroxypyrene-1,3,6-trisulfonic acid trisodium salt, HPTS) results in proton transfer to the formate of the EAF. The proton transfer predominantly occurs over picoseconds through a pre-formed hydrogen-bonded complex between HPTS and EAF. We investigate the longer range and longer timescale proton transport processes in the ionic liquid by obtaining the ground-state conjugate base ($RO^-$) dynamics from the congested transient-infrared spectra. The spectral kinetics indicate that the protons diffuse only a few solvent shells from the parent photoacid before recombining with $RO^-$. A kinetic isotope effect of near unity ($^{k_H}/_{k_D} \sim 1$) suggests vehicular transfer and transport of excess protons in this ionic liquid. Our findings provide a comprehensive insight into the complete photoprotolytic cycle of excess protons in a protic ionic liquid.

**TOC graphic**

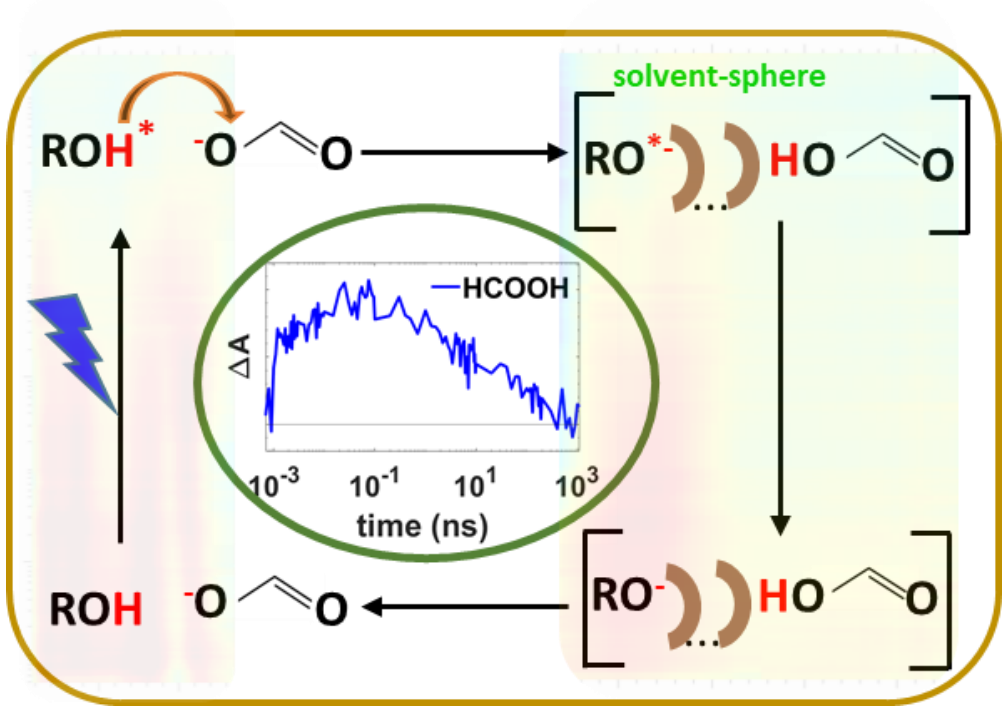

Protic ionic liquids (PILs) are interesting non-aqueous solvents because of their nonvolatility, thermal and electrochemical stability, and high ionic conductivity. These may in principle raise the operating temperature of a fuel cell to >120°C.[1-2] PILs are salts that are molten at room temperature and formed by reaction of a Brønsted-Lowry acid and a Brønsted-Lowry base. Mechanistic understanding of proton transport in PILs could help us to better realize their potential as proton conducting materials for practical applications such as electrolytes for hydrogen fuel cells.[3] Nevertheless, a great deal is unknown about the mechanisms of proton transport in PILs. Photoacids (ROH), where optical excitation leads to a transient increase in acidity (pH-jump), have been utilized extensively as a trigger to investigate the proton transfer processes through time-resolved spectroscopy.[4-23] Upon optical excitation, the acid dissociation constant, $K_a$, of 8-hydroxypyrene-1,3,6-trisulfonic acid trisodium salt (HPTS, Figure 1(a)) increases almost seven orders of magnitude (a p$K_a$ change from 7 to 0.4 in water),[20, 24] enabling the ultrafast release of protons into solution. This approach has provided valuable mechanistic insights into the proton transfer process in aqueous solvents for acid-base reactions. In the framework of the Eigen-Weller model, the bimolecular proton transfer reaction consists of a proton transfer step in a reactive complex followed by diffusive separation of products.[7, 17, 22, 25-26]

The transport of protons in ionic liquid media can be characterized by multiple approaches. Pulse-field gradient NMR techniques on imidazolium-based PILs can provide proton diffusion coefficients.[27-29] Time-resolved photoluminescence, which has a typical time window of 0.1-15 ns, provide excited-state proton transfer dynamics from photoacids to the anion in PILs.[30-31] Fuji et al. investigated the feasibility of proton transfer and the dynamics of associated intermediate states for naphthol based photoacids to PILs with different anionic basicity.[31] Recently, Thomaz et al. reported proton transfer dynamics from HPTS to an aprotic solvent 1-methylimidazole (an important cation for ionic liquids) through time-resolved photoluminescence spectroscopy.[32] Building upon the available literature,[7-8, 10-13, 19-20, 22, 30-32] in this work, we are able to probe the full photoprotolytic cycle (Figure 1(b)) of HPTS by widening the accessible spectroscopic timescale. This enables us to investigate short timescale proton transfer (picoseconds) and long timescale proton transport (nanoseconds-microseconds).

We investigated proton transfer and transport in the PIL ethylammonium formate (EAF, Figure 1(a)) from femtoseconds to milliseconds (~150 fs time resolution) using visible pump-infrared probe transient absorption measurements. The transient pump-probe measurements were performed in the time-resolved multiple probe spectroscopy (TRMPS) mode of operation, enabling measurement from femtoseconds-microseconds-milliseconds in a single experiment.[33] Observing the complete photoprotolytic cycle has enabled us to determine both the ultrafast steps of proton transfer from the photoacid to the PIL and the long-range proton transport process (Figure 1(b)). Viewed as a hydroxyl compound, ROH, photoexcitation of HPTS leads to a highly acidic excited state $ROH^*$, resulting in a prompt, reversible proton transfer to formate in EAF and generating a encounter pair (EP*).[12, 22] The EP* dissociates reversibly into individual components $RO^{*-}$ and formic acid (HCOOH) on the picosecond to nanosecond timescale. The $EP^*$ and $RO^{*-}$ also relax to the electronic ground state through radiative and nonradiative pathways on a few nanoseconds timescale.[7-8, 10-14, 17, 20, 22] The resulting ground state species, $RO^-$, then serves as a proton scavenger; it recombines with a proton to regenerate HPTS, completing the photoprotolytic cycle on the hundreds of nanoseconds timescale.

From our work, the key conclusions are as follows: (i) the proton transfer process predominantly proceeds through a 'tight-complex', where HPTS is H-bonded to formate prior to photoexcitation; (ii) the EP* is longer lived in EAF compared to water owing to the higher viscosity of EAF; and (iii) proton transfer and transport follow a vehicular mechanism with no significant kinetic isotope effect.

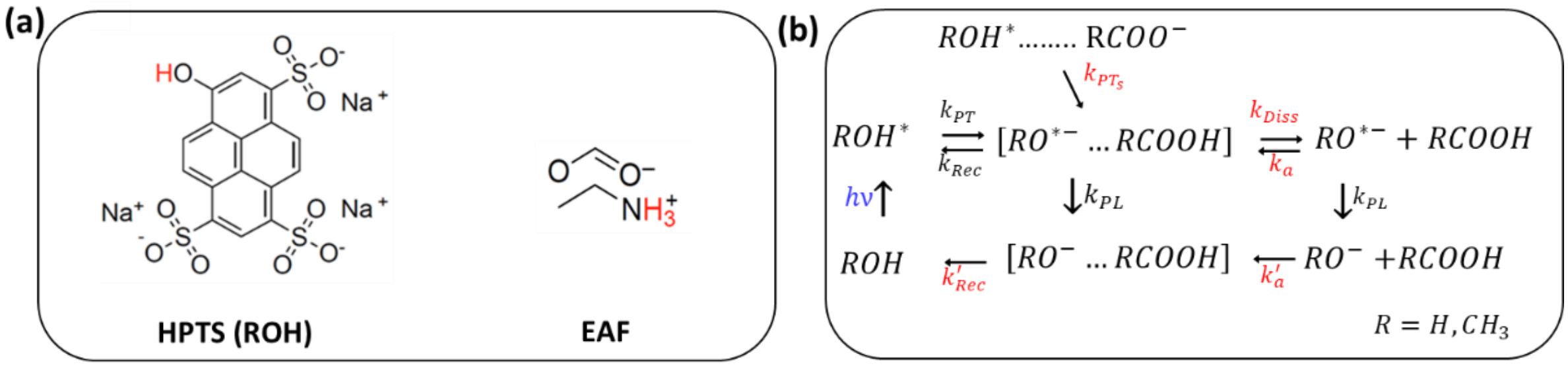

**Figure 1.** (a) Structure of 8-hydroxypyrene-1,3,6-trisulfonic acid trisodium salt (HPTS ≡ ROH) and ethylammonium formate (EAF). (b) A proposed photoprotolytic cycle of HPTS in water/acetate and EAF. Proton transfer from photoexcited HPTS (ROH*) to an acceptor (acetate or formate) can proceed reversibly *via* a direct hydrogen-bonded complex (termed 'tight-complex') to form an excited encounter pair (EP*) with forward and reverse reaction rate $k_{PT}$ and $k_{Rec}$, respectively.[7, 10-13, 19-20, 22] A fraction of proton transfer occurs at a rate an order of magnitude slower ($k_{PT_s}$) for species not initially hydrogen-bonded (termed 'loose-complex').[7, 10-13, 19-20, 22] The EP* can dissociate ($k_{Diss}$) into individual constituents ($RO^{*-}$ and HCOOH) or generate ($k_{PL}$) a ground state EP through photoluminescence. The conjugate base $RO^{*-}$ can re-form the EP* bi-molecularly with HCOOH ($k_a$) or become de-excited ($k_{PL}$) to $RO^-$ through photoluminescence. The $RO^-$ and HCOOH bi-molecularly produce the ground-state EP (at rate $k'_a$), whereupon proton transfer ($k'_{Rec}$) from HCOOH to $RO^-$ regenerates the HPTS, completing the cycle.

20 mM HPTS dissolved in EAF was photoexcited (~10 μm pathlength) with a 400 nm laser pulse (~150 fs) and the resulting changes in the absorbance ($\Delta A = A_{pumped} - A_{unpumped}$, $A$ = absorbance) were monitored in the mid-infrared region. The EAF synthesis and characterization are described in section 1, SI and Figure S1, SI. The transient-infrared experiment has been described in detail elsewhere.[33] A synchronised pump (400 nm, 1 μJ/pulse) and probe (1400–1900 cm$^{-1}$, <0.1 μJ/pulse) operated at 1 and 10 kHz, respectively. The sample was rastered in the focal plane to avoid degradation of HPTS. Figure 2(a) shows the transient-infrared spectra *vs* time 2D-colormap for photoexcitation of HPTS in EAF at 400 nm.

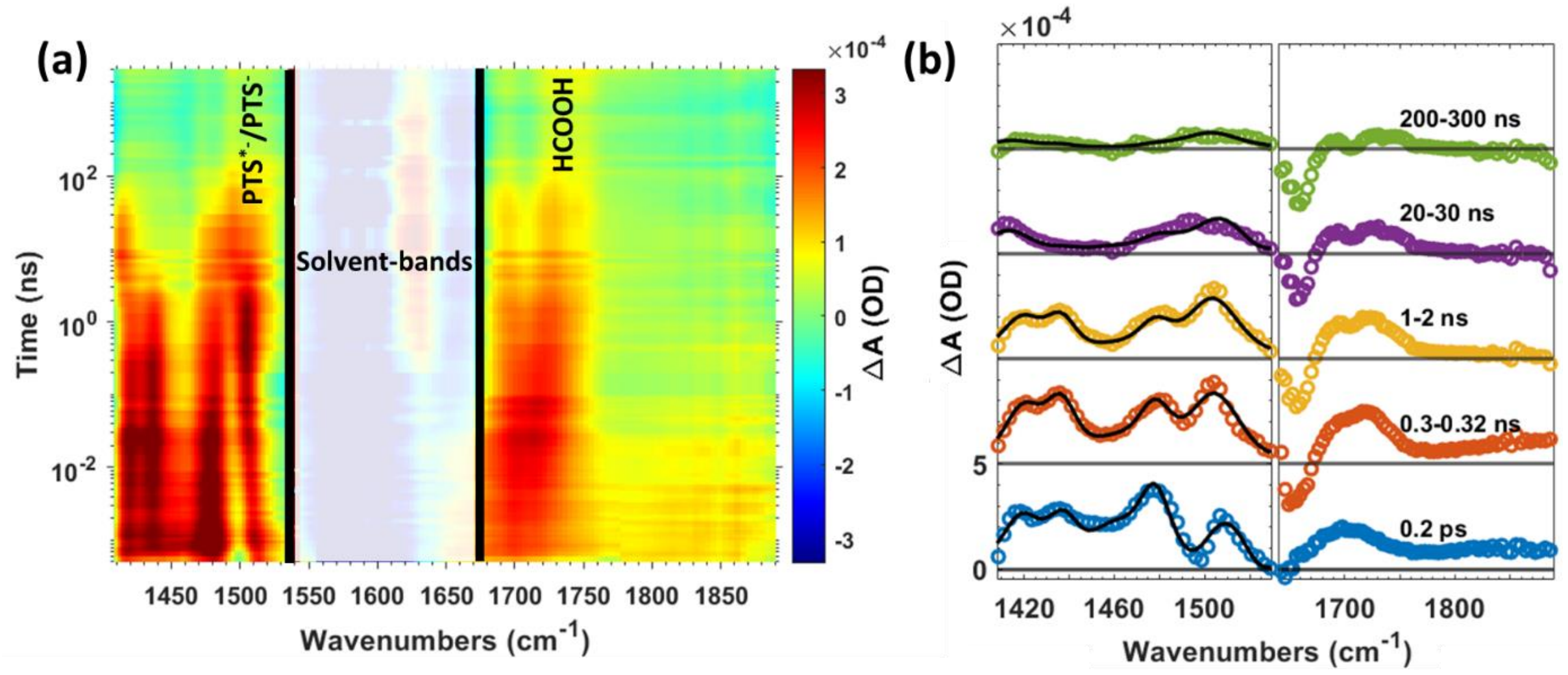

**Figure 2.** (a) 2D colormap representing the transient absorbance ($\Delta A$) of 20 mM HPTS dissolved in ethylammonium formate (EAF) upon 400 nm pump excitation (fwhm ~150 fs). (b) Transient spectra at representative pump-probe delay times. The black lines are spectral lineshape fits of the transient spectra, as discussed in the text. The apparent splitting of the formic acid band (~1710 cm$^{-1}$) accompanies the loss of excited-state species and results from thermal shifts of the adjacent, strong EAF absorption band (Figure S2, SI).

The transient-infrared spectra provide marker modes for the key species in the photoprotolytic cycle (Figure 2). The 1530-1650 cm$^{-1}$ region is excluded from the analysis because strong infrared absorption from the formate carbonyl ($-C = O$) groups mask the transient-infrared signal. A broad photoinduced absorption band at ~1725 cm$^{-1}$ appears due to proton transfer from photoexcited HPTS (ROH$^*$) to formate, creating formic acid.[7, 10, 12-13, 22] This feature follows the growth of the RO$^{*-}$ photoinduced absorption ~ 1435 cm$^{-1}$ and ~1503 cm$^{-1}$. The formic acid absorption at ~1725 cm$^{-1}$ is free from overlap with other infrared bands at all pump-probe delay times, allowing the spectral amplitude of this species to be obtained simply by averaging the transient absorption band in the region 1720-1735 cm$^{-1}$.

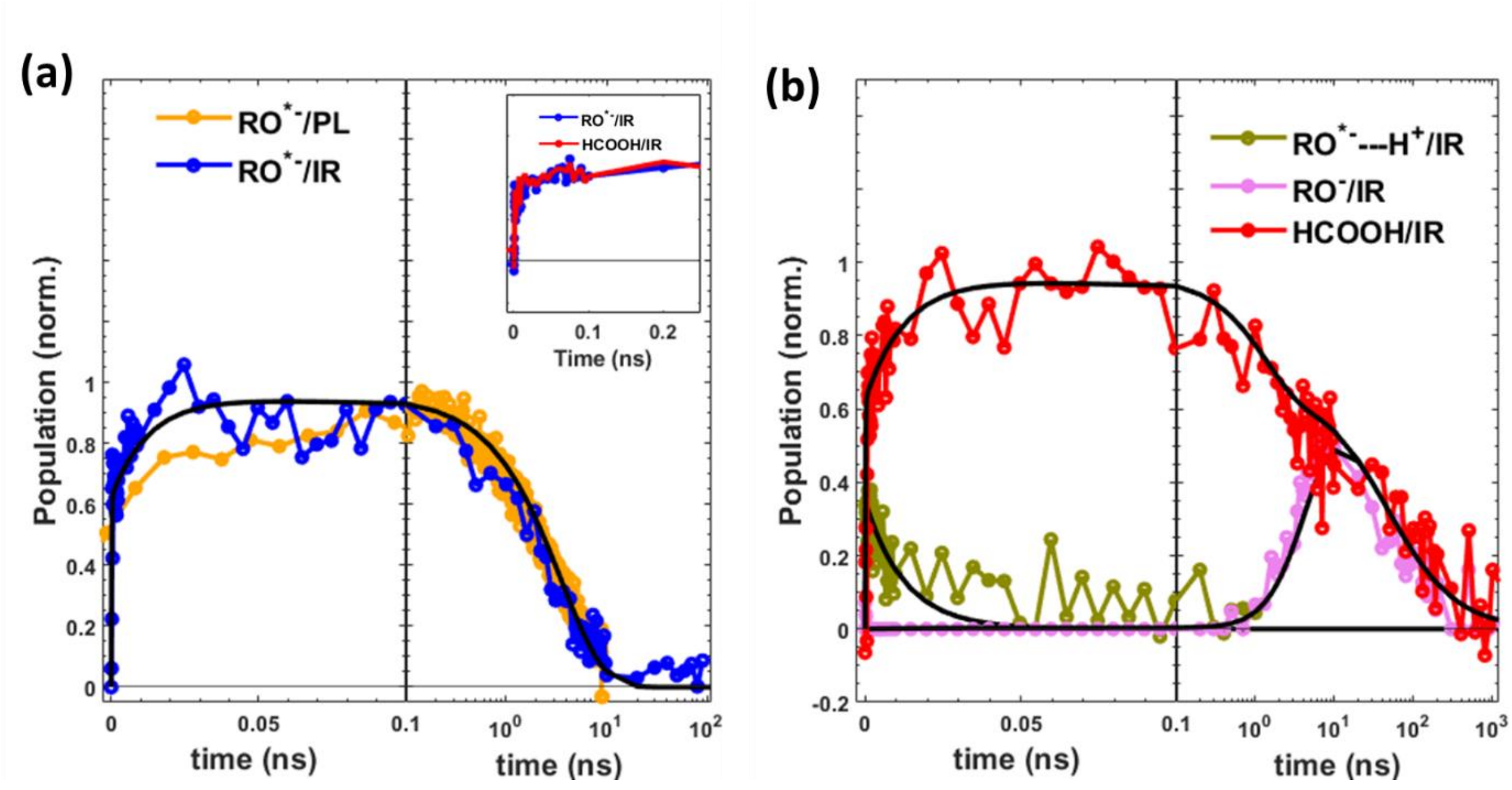

**Figure 3.** (a) Transient-infrared kinetics of RO$^{*-}$ (blue) plotted along with the photoluminescence decay (orange). (b) Kinetics of loosely bound proton of ROH$^*$ (olive, average in the 1770-1850 cm$^{-1}$), RO$^-$ (pink) and formic acid (red, average in the region 1720-1735 cm$^{-1}$). For RO$^{*-}$ and RO$^-$ the kinetics are obtained from spectral model fitting of transient infrared absorption data (details in text). The black lines represent fits according to the kinetic

model of Figure 1(b). The inset in (a) shows early time comparisons of kinetics between HCOOH (1720-1735 $cm^{-1}$) and $RO^{*-}$ (1501-1505 $cm^{-1}$).

To isolate the kinetics of the involved species from the congested probe spectra, we fit the transient-infrared spectra in the region 1425 – 1530 $cm^{-1}$ (primarily aromatic ring vibrational modes), with models for the infrared spectra of ROH, $ROH^*$, $RO^{*-}$ and $RO^-$, each comprising a sum of Voigt lineshapes[34] (Section 2, SI). The resulting 4-component spectra (black lines) are overlaid with the raw transient spectra signal amplitudes at selected frequencies, showing good agreement (Figure 2(b)). The models for the infrared spectra of these three species are determined by analysis of the evolution of the transient spectra and by making use of spectra available in the literature (Figure S3, SI).[11-13, 22] Singular value decomposition (SVD) failed to provide chemically meaningful components due to significant overlap in frequency and time of the spectra of $ROH^*$, $RO^{*-}$ and $RO^-$, especially $RO^{*-}$ and $RO^-$.

Figure 3 (a-b) shows the kinetics of the formic acid (HCOOH) band and the extracted kinetics of $RO^{*-}$ and $RO^-$. The formic acid shows bimodal growth comprising a fast pulse-width limited component and an order of magnitude slower component growing over ~100 ps. Similar to HPTS/acetate studies in water, we attribute the fast-growth component to instantaneous (pulse-width limited) proton transfer from photo-excited HPTS ($ROH^*$) to formate in hydrogen-bonded tight complexes pre-existing prior to photoexcitation.[7, 22] The slower growth component can be ascribed to $RO^{*-}$–formate pairs which are not strongly hydrogen-bonded (weakly-complexed) at time of photo-excitation and termed as 'loose-complex'. These must reorganize to form an encounter pair prior to proton transfer. The formic acid and $RO^{*-}$ bands have identical growth dynamics (Figure 3(a), inset) suggesting direct proton transfer occurs both in the tight and the loose complexes. The growth of $RO^-$ follows the decay of $RO^{*-}$ (Figure 3). The $RO^-$ decays by accepting a proton from formic acid to regenerate HPTS (ROH) and thus both $RO^-$ and formic acid signals decay identically (Figure 3(b)). We observe a broad photoinduced absorption at frequencies blue of ~1760 $cm^{-1}$ in the transient-infrared spectra (Figure 2). In the case of HPTS in water, this has been attributed to loosely bound protons of $ROH^*$.[12-13] Upon photoexcitation in water, the O-H bond in HPTS weakens and the hydrogen bond with solvent water strengthens. The broad infrared absorption results from this loosely bound, highly polarizable proton. Based on this explanation, the broadband infrared absorption feature in HPTS/EAF likely results from similarly loosely bound protons of $ROH^*$ in the hydrogen bonding environment of formate.

To corroborate our model, we compared the kinetics of $RO^{*-}$ obtained from the spectral model fitting to the photoluminescence decay (Figure 3(a)) of $RO^{*-}$ around its emission maxima of 510 nm (Figure S1, SI). The excellent agreement between the transient-infrared and photoluminescence measurements supports the accuracy of the infrared spectral lineshape fits and interpretation.

The chemical kinetics scheme (Figure 1(b)) reproduces the observed population dynamics and provides quantitative estimates of proton transfer rates, ion-pair separation, and proton recombination (Table 1). A simultaneous fit of the loosely-bound proton, $RO^{*-}$, $RO^-$ and formic acid kinetics through the non-linear least square method was used to yield best-fit parameters (Section 3, SI). The populations obtained from the kinetic model (Figure 3, black lines) agree well with experimental data from picoseconds to microseconds. The bimodal formic acid and $RO^{*-}$ growth is best estimated with a pulse-width limited rise, $k_{PT} \approx$ 150 fs, followed by a slower growth component, $k_{PT_s}$ , of 113.4 $ns^{-1}$ ($1/k_{PT_s}$ = 8.8 ps). The decay of the loosely-complexed fraction of ROH*, $k_{PT_s}$ as indicated by the dynamics of broadband infrared feature (1760-1850 cm$^{-1}$), correlates with the slow growth component of formic acid (Figure 2(b)). Thus, after photoexcitation, the proton becomes loosely bound in ROH* but does not instantly transfer to formate, as in the direct complex. Structural rearrangements of the ROH* and EAF over longer timescales (tens of picoseconds) are required for the proton to transfer. About 90% of the photogenerated formic acid decays back to formate in ~300 ns.

**Table 1.** Estimated rate constants based on the kinetic model in Figure 1(b) for HPTS/EAF and DPTS/deuterated EAF (EAF-3D).

| Rate constants [a] | HPTS/EAF | DPTS/EAF-3D | $k_H/k_D$ |
|---|---|---|---|
| $k_{PT}$ [b] | (150 fs)$^{-1}$ | (150 fs)$^{-1}$ | - |
| $k_{PT_s}$ | $113.4 \pm 1.2\ ns^{-1}$ | $112.4 \pm 0.98\ ns^{-1}$ | 1.01 |
| $k_{Rec}$ | $396.5 \pm 3.4\ ns^{-1}$ | $401.4 \pm 2.9\ ns^{-1}$ | 0.99 |
| $k_{Diss}$ | $0.40 \pm 0.01\ ns^{-1}$ | $0.39 \pm 0.01\ ns^{-1}$ | 1.03 |
| $k_{PL}$ [c] | $0.30 \pm 0.01\ ns^{-1}$ | $0.29 \pm 0.01\ ns^{-1}$ | 1.03 |
| $k_a$ [d] | $0.5 \times 10^{10}$ M$^{-1}$s$^{-1}$ | $0.5 \times 10^{10}$ M$^{-1}$s$^{-1}$ | - |
| $k'_a$ | $(3.27 \pm 0.36) \times 10^{10}$ M$^{-1}$s$^{-1}$ | $(3.02 \pm 0.24) \times 10^{10}$ M$^{-1}$s$^{-1}$ | 1.08 |
| $k'_{Rec}$ | $164.1 \pm 1.5\ ns^{-1}$ | $161.3 \pm 1.3\ ns^{-1}$ | 1.02 |

[a]The rates were optimized by taking reported values in literature as a guide.[7-8, 10-13, 20, 22, 35-37] The uncertainties represent the standard errors of the estimated parameters. [b] Faster than the time-resolution of measurement. [c]$k_{PL}$ was optimized close to the photoluminescence lifetime. [d] As $k_a$ is order of magnitude smaller[19] than $k_{Diss}$ a constant rate was used for better fit.

In order to determine the mechanism of proton transport, we determined the kinetic isotope effect ($KIE = {}^{k_H}/_{k_D}$) on the fitted rates through H-D isotope substitution. For HPTS in water, KIEs ${}^{k_H}/_{k_D} \sim 1.4 \sim \sqrt{{}^{m_D}/_{m_H}}$ have been reported previously for rates in the proton transfer cycle, indicating that steps in the cycle (marked red in Figure 1(b)) depend on proton mass.[10, 12-13, 20] This finding has been used to identify steps where Grotthuss proton transport occurs, [38] for which protons hop from acceptor to acceptor. We examined DPTS/EAF (3D), where all the exchangeable H were replaced with D, to record analogous transient absorption spectra (Figure S4(a), SI).

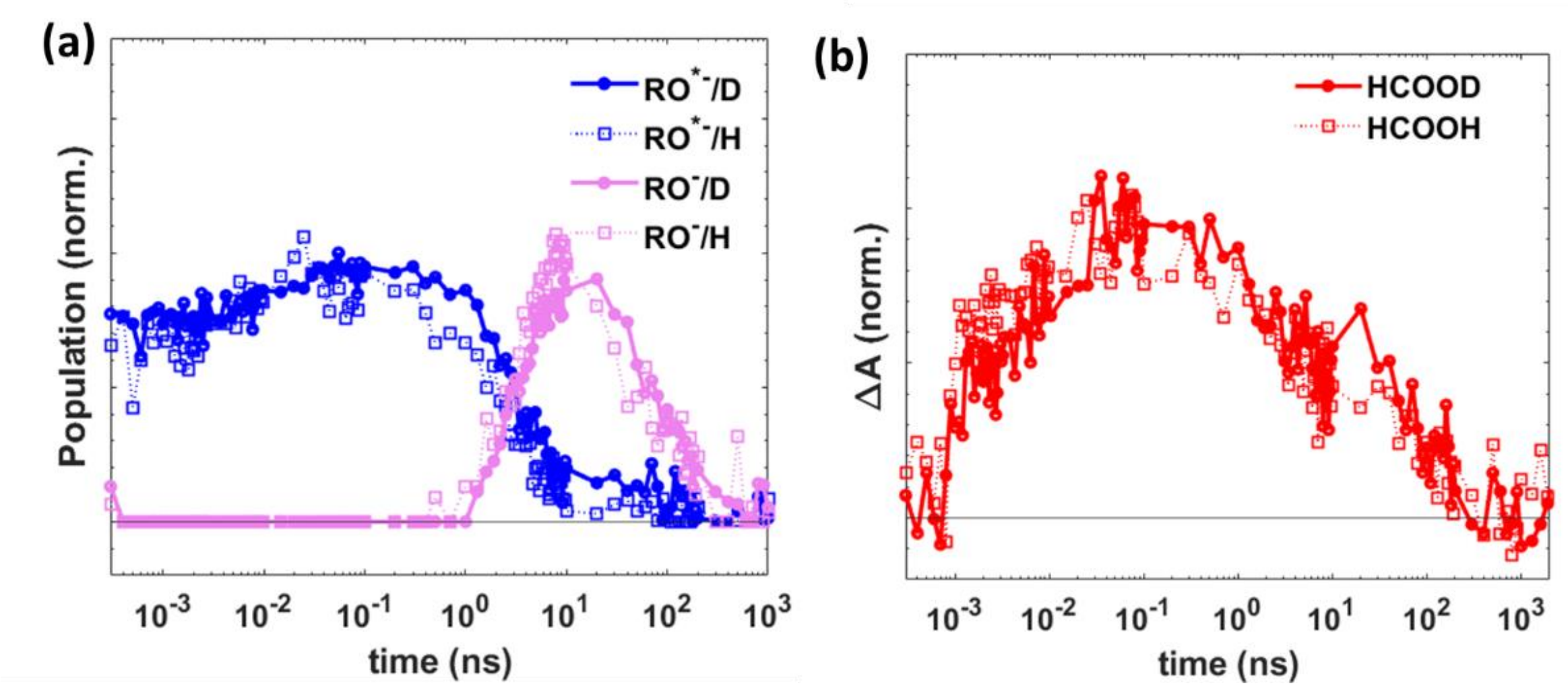

**Figure 4.** Comparison of the kinetics of (a) $RO^{*-}$ and $RO^{-}$ and (b) formic acid in HPTS/EAF (dotted line) and HPTS/EAF-3D (solid line). The populations of $RO^{*-}$ and $RO^{-}$ were obtained from spectral model fitting, whereas the population of formic acid was obtained from the average transient absorption between 1720 and 1735 $cm^{-1}$.

The kinetics of protonated and deuterated samples are nearly identical (Figure 4). Figure S4(b), SI shows the spectral line shape-fit plots of the transient data and Figure S5, SI depicts the kinetic decay along with fits according to the kinetic model in Figure 1(a), with best-fit parameters (Table 1). The observed $^{k_H}/_{k_D}$ is less than 1.1 for all rate constants (Table 1), suggesting vehicular proton transfer and transport at all stages of the photoprotolytic cycle.

How far does the photogenerated proton travel in EAF? We can estimate the excess proton diffusion length ($L_D = \sqrt{D\tau}$) by taking into account the combined diffusion coefficient ($D = D_{HCOOH} + D_{RO^-}$) for formic acid and $RO^-$ in EAF, and the lifetime ($\tau \sim 300$ ns) of the transient formic acid. The estimated diffusion length is ~5 nm (Section 4, Supporting Information). This distance corresponds to about ~8-10 solvation shells, where one solvation shell in EAF is $\sim 0.5$ nm, as determined by neutron diffraction.[39]

To understand the differences of EAF to aqueous systems, we compared the rates obtained for HPTS/EAF to equivalent measurements of HPTS/acetate (1M) in $D_2O$. The kinetics of the different species for the aqueous system were again obtained through spectral lineshape fitting and kinetic modelling (Figure S6, SI) to obtain the rate constants (Table S1, SI). [7-8, 10-13, 20, 22, 35-37, 40] The proportion of the tight complex is higher in EAF (~70%) than water/acetate (~40%) because the acceptor base (formate) is also the solvent in EAF and higher in concentration (~13M). Nevertheless, the $RO^{*-}$ growth (and consequently the acetic/formic acid growth) rate constant is similar in water/acetate and EAF.The decay of EP* was ~9 times slower in EAF than in water which can be attributed to the higher diffusion coefficient of acetic acid in water compared with formic acid in EAF. EAF's higher viscosity (23.1 mPa s)[41] than water (1 mPa s) likely leads to this slower EP* decay. $RO^-$ is longer lived in water, as the bimolecular association rate ($k'_a$) of $RO^-$ and acetic acid to form the EP in water (1M acetate) is about two times slower than the association of $RO^-$ with formic acid in EAF (Table 1). The higher diffusion coefficient in water and Grotthuss transport can result in a larger separation between the acid and $RO^-$ making the $k'_a$ smaller in water.

Previous studies on proton diffusion in protic ionic liquids using pulsed-field gradient NMR have shown vehicular proton transport in equimolar (1:1) protic ionic liquids based on imidazolium cations and bis(trifluoromethylsulfonyl)imide anions.[27-29] On the other hand, Grotthuss transport[38] would be highly desirable in PIL-based fuel cells. Lin et al. have shown evidence that adding water (6 volume %) enhances the proton conductivity through Grotthuss transport in PILs with highly acidic cations ($pK_a$~0) such as 2-sulfoethylmethylammonium triflate [2-Sema][TfO].[42] On the other hand, Grotthuss transport as a proton transport mechanism has been demonstrated in pseudo ionic liquids (equimolecular mixtures of N-methylimidazole and acetic acid).[43-44] Recently, Ingenmey *et al.* have theoretically predicted combinations of substituted cation and anions for pseudo protic ionic liquids that may display Grotthuss transport.[45] Transient infrared spectroscopy will be an important tool for deciphering the proton transport mechanisms in these new PILs.

In conclusion, we have investigated proton transfer and long-range proton transport from the excited state of a photoacid HPTS in a protic ionic liquid ethylammonium formate through pump-probe vibrational spectroscopy. The proton transfer predominantly proceeds through hydrogen-bonded HPTS-EAF complexes having ultrafast proton transfer rate (<150 fs) whereas proton transfer in a proportion of loosely-complexed pairs is an order of magnitude slower. The long-range proton transport was deciphered by analysing the kinetics of both excited and ground-state species with transient-infrared and photoluminescence, suggesting proton transport to a few (~10) solvent shells in the protic ionic liquid. The absence of the kinetic isotope effect suggests the presence of vehicular transfer and transport of excess protons in EAF across the photoprotolytic cycle. The wide time-range study presented here improves on transient-infrared spectroscopic approaches for studying the solvent influence on whole photoprotolytic cycles, paving the way to investigate long-range excess proton transport in further PIL systems of interest, in situ and in operando.

**Acknowledgements**

We thank Prof Stanley Botchway for help with the photoluminescence measurements. We also thank Prof Tony Parker for help with the experiments. We are grateful for funded access (App16230054) to the ULTRA labs of the Central Laser Facility (CLF), STFC. This work was supported by STFC and by a UKRI Future Leaders Fellowship grant (MR/S015574/1). We are grateful to the Research Complex at Harwell (RCaH) for access to experimental facilities. The authors acknowledge financial support from the National Science Foundation (USA) (CHE-1454105 and CHE-1954848).

**Associated Content: Supporting Information**

Sample preparation and characterization, Spectral lineshape fitting and kinetic modelling, Transient data for DPTS/EAF-3D and HPTS/acetate (1M) in $D_2O$.

**Conflicts of interest**

The authors declare no conflict of interest.

# Supporting Information:

# pH-jumps in a Protic Ionic Liquid Proceed by Vehicular Proton Transport

Sourav Maiti[§], Sunayana Mitra[¶], Clinton A. Johnson[¶,$], Kai C. Gronborg[¶], Sean Garrett-Roe[¶*] and Paul M. Donaldson[§*]

[§]*Central Laser Facility, RCaH, STFC-Rutherford Appleton Laboratory, Harwell Science and Innovation Campus, Didcot, United Kingdom*

[¶]*Department of Chemistry, University of Pittsburgh, 219 Parkman Avenue, Pittsburgh, Pennsylvania 15260, USA*

*Corresponding Authors

Sean Garrett-Roe: sgr@pitt.edu

Paul M. Donaldson: paul.donaldson@stfc.ac.uk

[$]Current address: *Department of Chemistry, Davis and Elkins College, 100 Campus Dr., Elkins, WV 26241, USA*

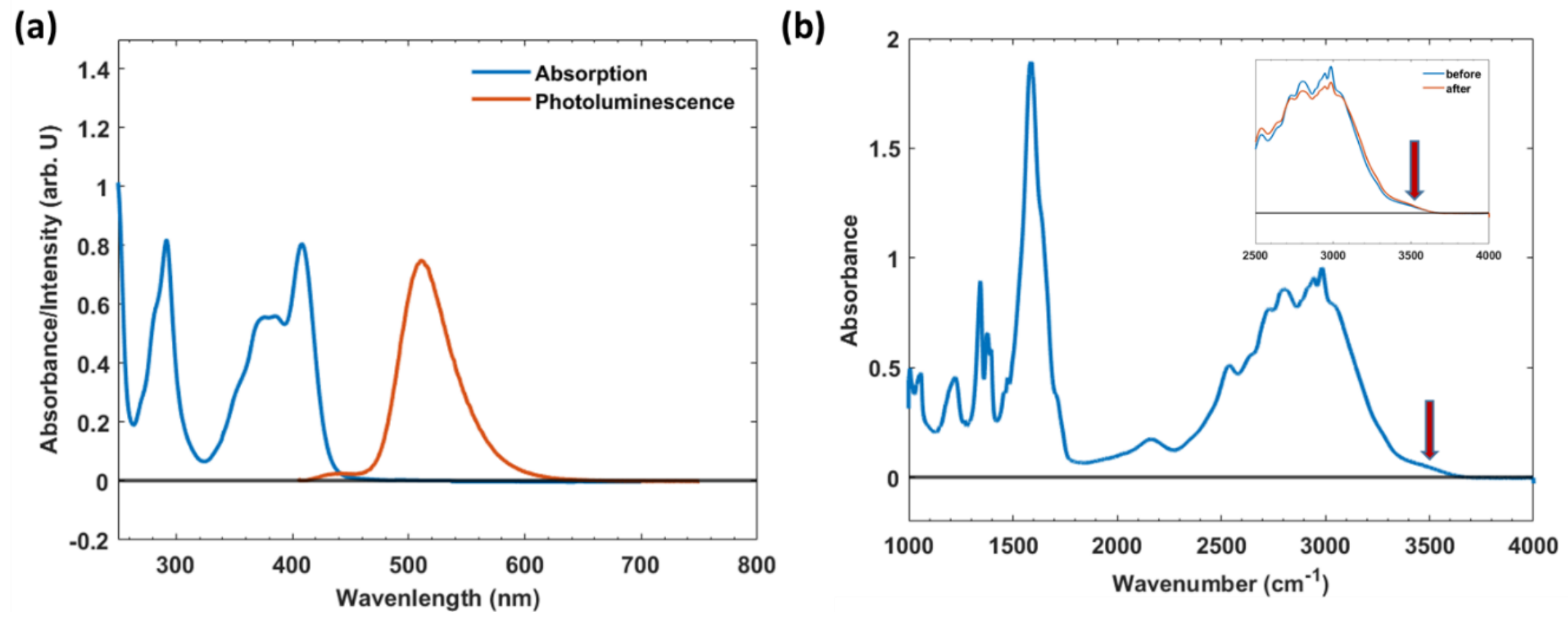

**Figure S1.** (a) UV-vis absorption and photoluminescence (400 nm excitation) spectra of HPTS/EAF. The photoluminescence spectrum shows $RO^{*-}$ emission at ~ 510 nm is predominant. (b) FTIR spectrum of EAF/HPTS. The arrow ~3500 $cm^{-1}$ shows small (< 1 wt %) water content. (Inset) FTIR before and after TRMPS experiment showing water content remains the same.

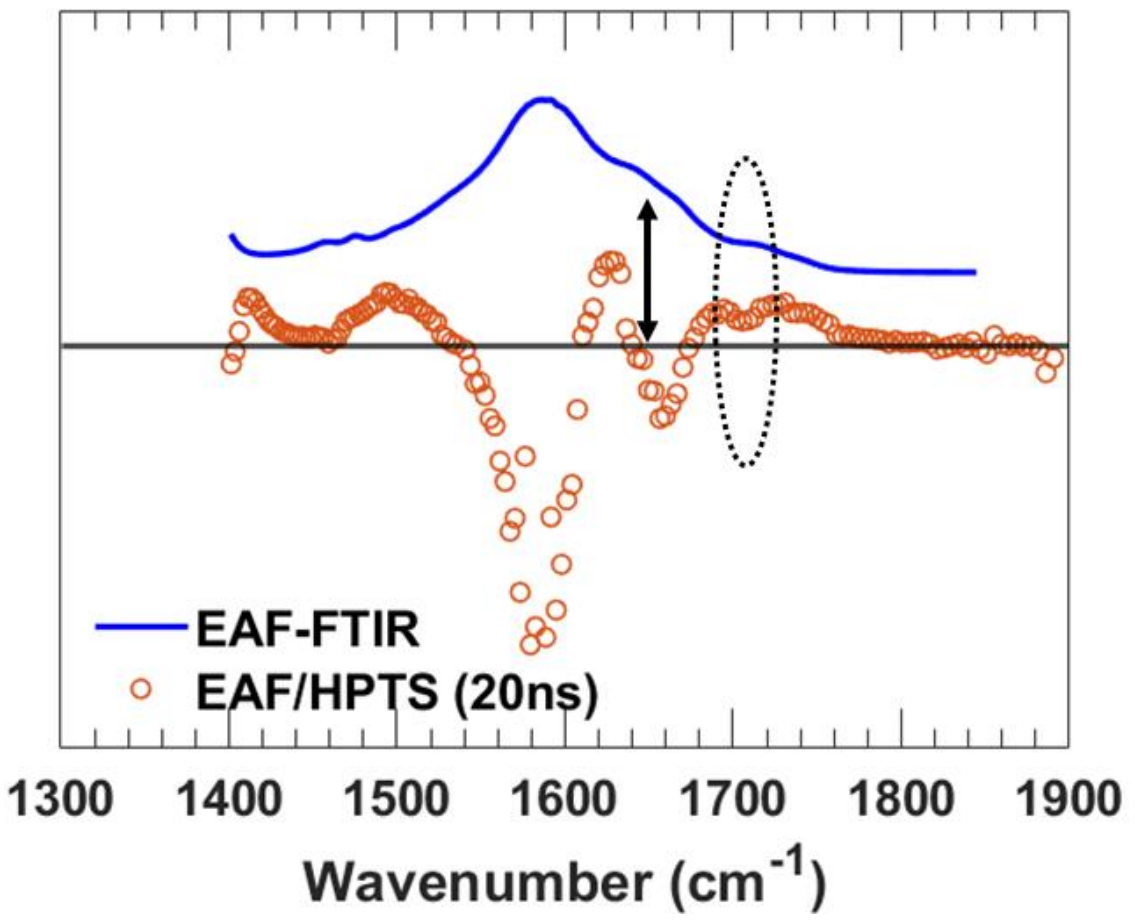

**Figure S2.** An EAF FTIR spectrum and a transient absorption spectrum of HPTS/EAF at 20 ns delay. The apparent split around 1710 $cm^{-1}$ in the formic acid band is due to an EAF band around the same region.

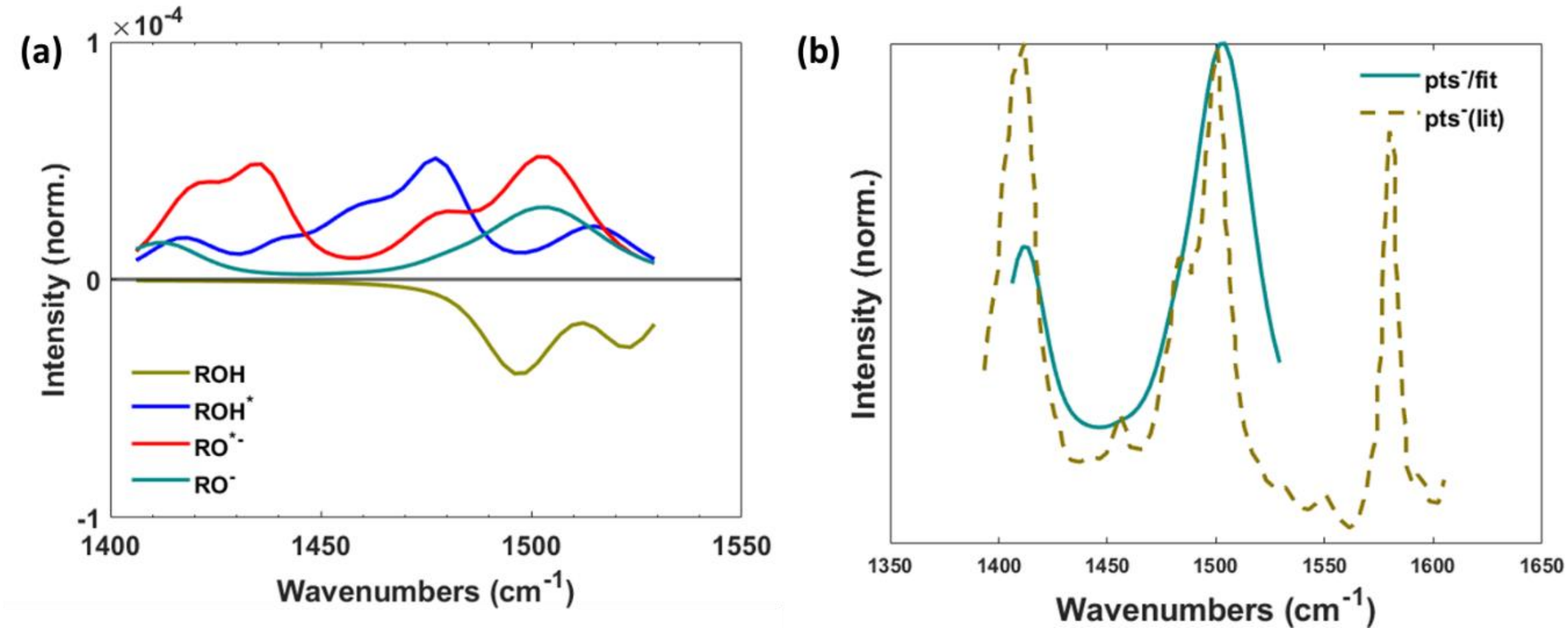

**Figure S3.** (a) The spectra of ROH*, ROH*- and ROH- as determined from Voigt profile analysis. (b) Comparison of Voigt spectra of PTS- with literature[1].

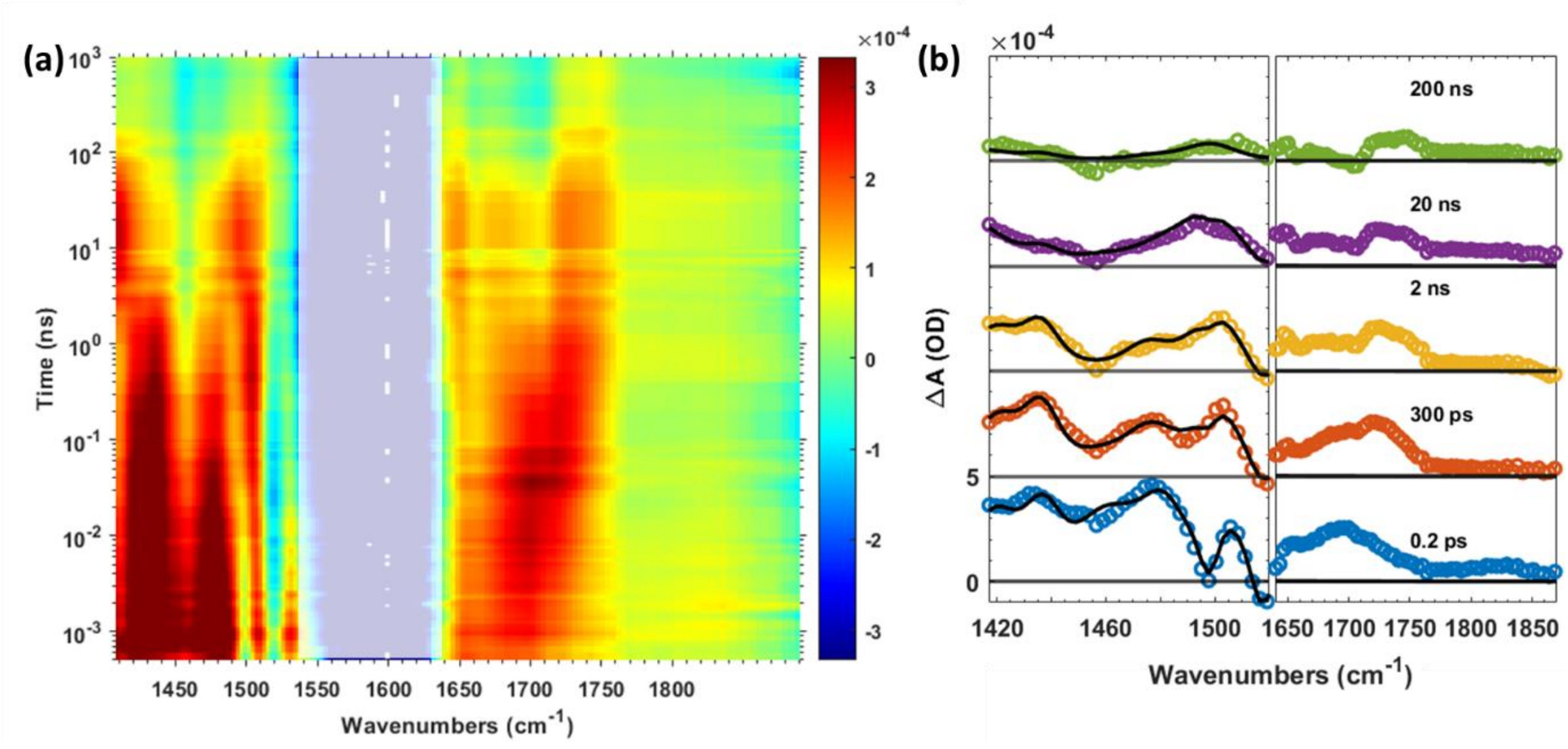

**Figure S4.** (a) 2D colourmap representing the transient absorbance (ΔA) of 20 mM DPTS dissolved in deuterated ethylammonium formate (EAF-3D) upon 400 nm pump excitation (fwhm ~150 fs). (b) Fitting the transient spectra (data points) with Voigt profile (solid line) of ROH, ROH*, ROH*- and ROH- at representative delay times in DPTS/EAF-3D.

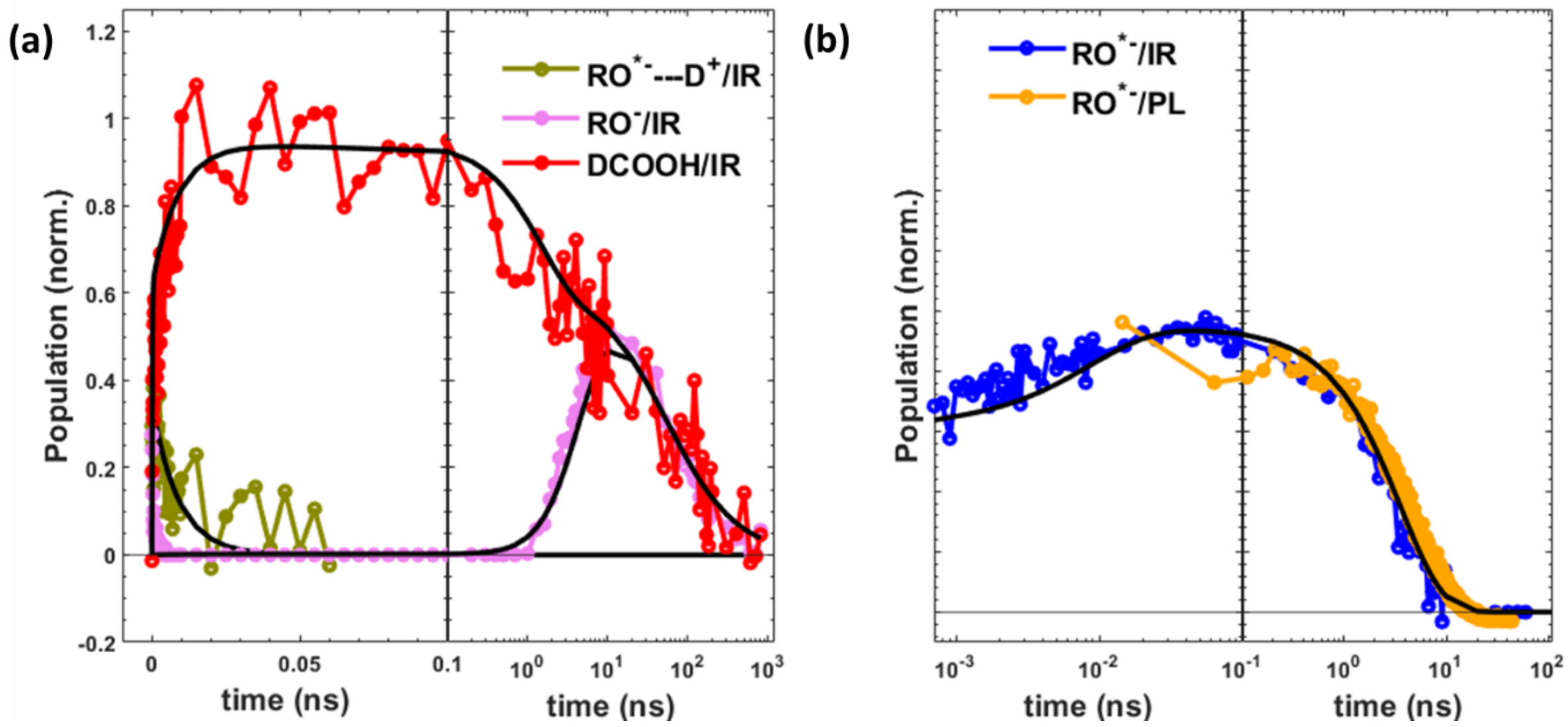

**Figure S5.** (a) Kinetics of RO*- and the photoluminescence (PL) decay in DPTS/EAF-3D. (b) Kinetics of loosely bound protons (from an average of signal across 1770-1850 $cm^{-1}$), $RO^-$ and formic acid (average across the region 1720-1735 $cm^{-1}$). The kinetics for $RO^{*-}$ and $RO^-$ are obtained from spectral model fitting of transient infrared absorption data. The black lines represent fits according to the kinetic model in Figure 1.

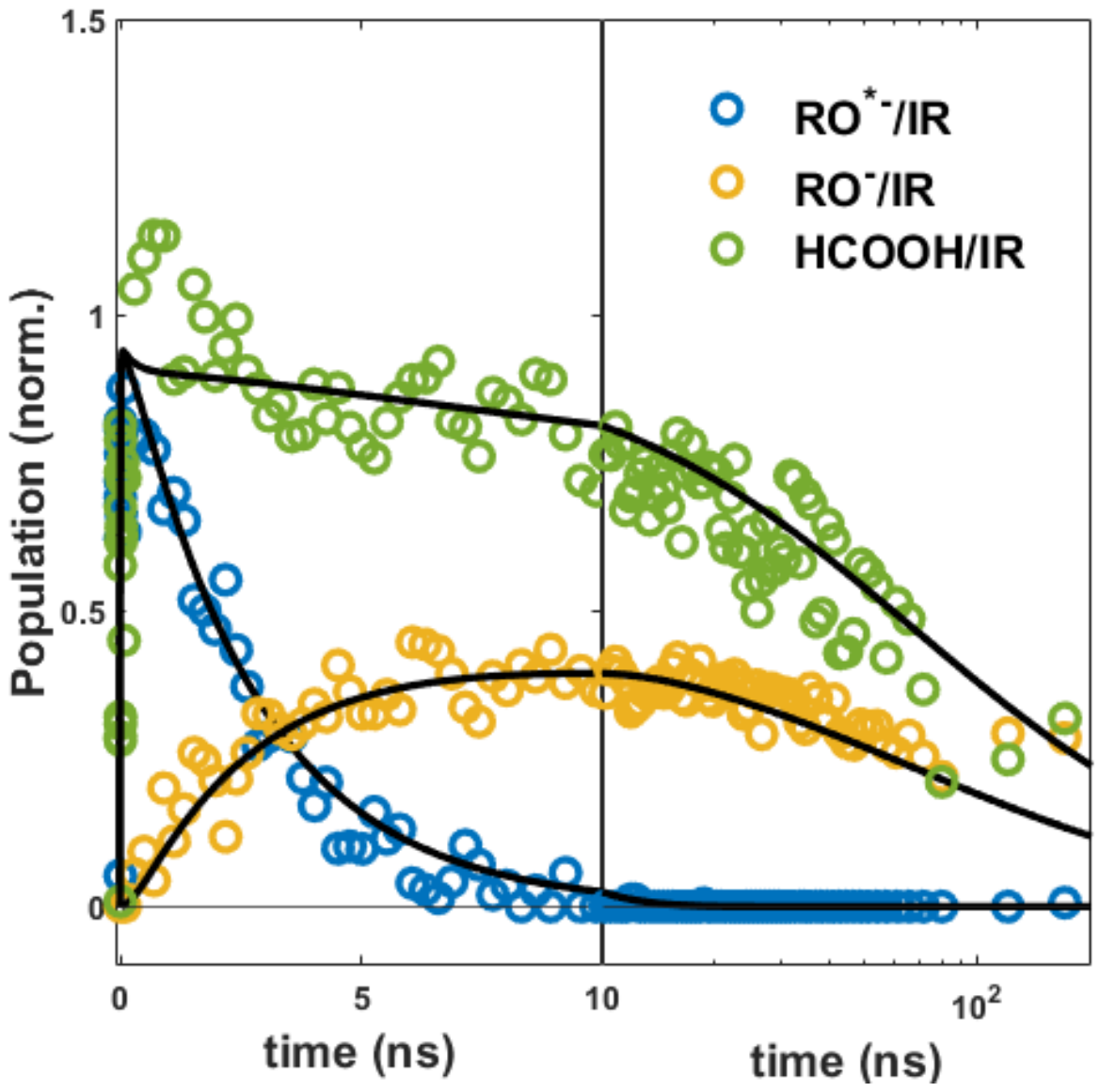

**Figure S6.** Extracted kinetics of $RO^{*-}$ and $RO^{-}$ from the spectral model fitting of transient data and kinetics of acetic acid for HPTS/1M sodium acetate/$D_2O$. Black lines are the fits according to the kinetic model (scheme 1 and section 3, SI).

**Table S1.** Fitted rate constants for HPTS/1M sodium acetate/$D_2O$.[a]

| Rate constants | HPTS/acetate (1M)/$D_2O$ |
|---|---|
| $k_{PT}$ | $(150\ fs)^{-1}$ |
| $k_{PT_s}$ | $113.2 \pm 6.1\ ns^{-1}$ |
| $k_{Rec}$ | $399.9 \pm 21.0\ ns^{-1}$ |
| $k_{Diss}$ | $3.4 \pm 0.2\ ns^{-1}$ |
| $k_a$[b] | $5 \times 10^{10}$ $M^{-1}s^{-1}$ |
| $k_{PL}$ | $0.37 \pm .01\ ns^{-1}$ |
| $k'_a$ | $(1.57 \pm 0.10) \times 10^{10}$ $M^{-1}s^{-1}$ |
| $k'_{Rec}$ | $150.1 \pm 34.8\ ps^{-1}$ |

[a] Rate constants are optimized within acceptable values as reported in the literature.[1-11] [b] Fixed rate constant taken from ref. [9] was used.

## 1. Experimental section

### a. Chemicals:

HPTS (Sigma Aldrich, $\geq 97\%$), Ethylamine (70 %wt in water, Sigma Aldrich), Formic acid (Sigma Aldrich), sodium acetate (Sigma Aldrich), $D_2O$ (99.9 atom%, Sigma Aldrich) were used as received.

### b. Preparation of samples

EAF was synthesized by titrating ethylamine with formic acid in an equimolar ratio at -70°C. The synthesized EAF was Schlenk vacuum line dried for seven hours to reduce the water content below 1 wt%. The EAF formation was confirmed through FTIR and $^1$H-NMR. The EAF was dried a second time prior to performing the TRMPS experiments.

Deuteration was performed by placing the dry, newly synthesized EAF with 5 times excess of $D_2O$ in a flask. The solution was held at ~35 °C and stirred for ~12 hours to ensure full exchange. A vacuum pump setup removed the heavy water in the solution. This deuteration process was repeated three times to ensure full labile H to D exchange, giving the desired EAF-3D. FTIR and $^1$H NMR spectroscopies characterized the new compounds. In FTIR spectroscopy, following the ND/NH stretch peak ratio throughout the exchange shows an increasing ND band (~2200 cm$^{-1}$) intensity to over 98% of the NH band (~3000 cm$^{-1}$). NMR spectroscopy was used to monitor the disappearance of the $NH_3$ peak to ensure complete H/D exchange. At the end of the exchange, the PILs were again dried at 65-70 °C using a vacuum pump setup, and the decreasing OD stretch of $D_2O$ (~2500 cm$^{-1}$) intensity in an FTIR spectrometer was checked.

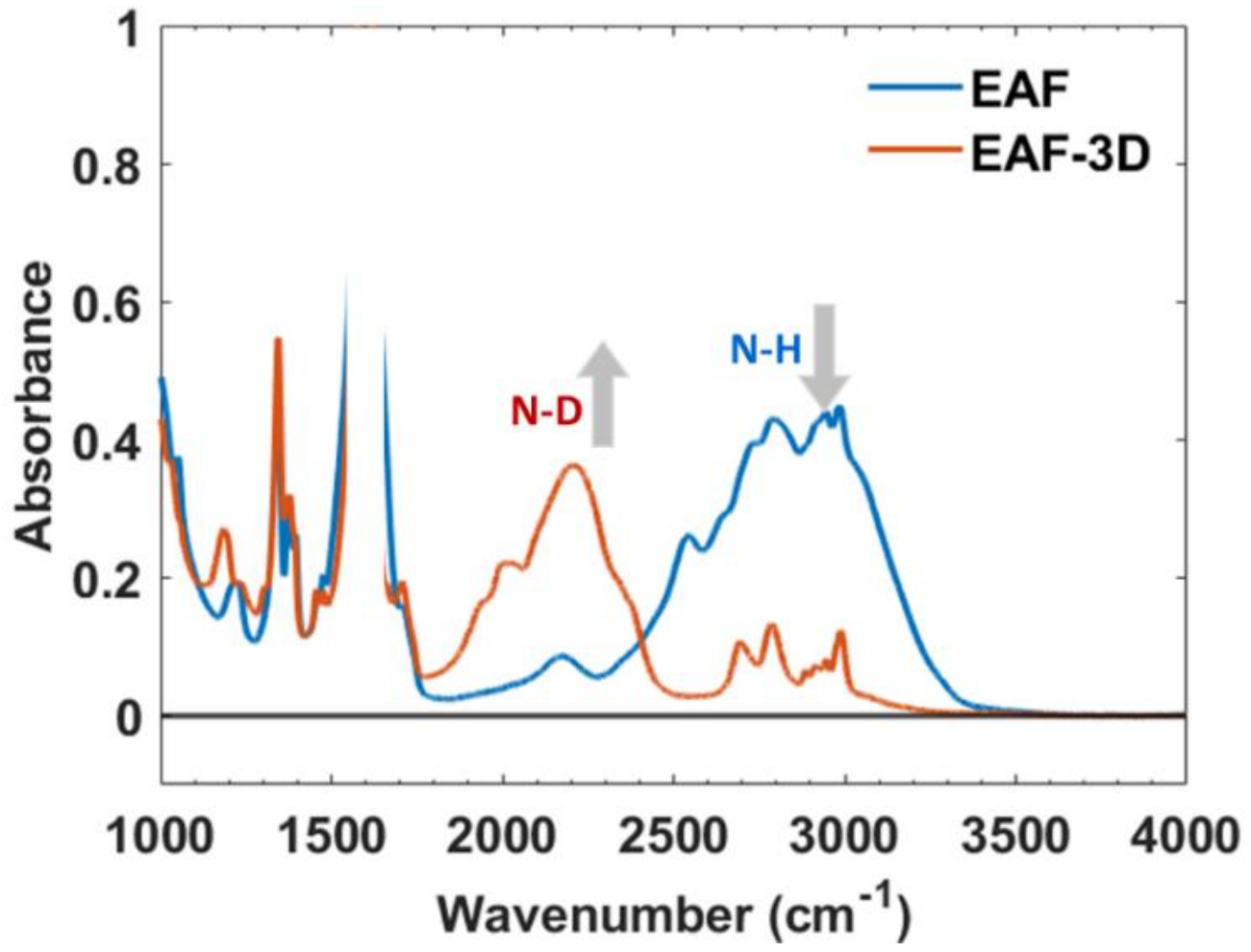

**Figure S7.** FTIR spectra of EAF and EAF-3D.

The FTIR spectrum in Figure S7 shows the appearance of an N-D stretching band at the expense of the N-H stretching band, confirming the H/D exchange. The around region ~1570 $cm^{-1}$ is excluded from the plot due to saturation of the acid carbonyl band.

20 mM solutions of HPTS in EAF were prepared in a glovebag filled with dry $N_2$. For aqueous experiments, 20mM of HPTS was dissolved in $D_2O$ with 1M sodium acetate. For FTIR and TRMPS measurements, the samples were placed in a Harrick cell with $CaF_2$ windows and no spacer. The Harrick cell was assembled inside a dry $N_2$ filled glovebag.

### c. Photoluminescence lifetime measurements:

Photoluminescence lifetimes were obtained using the time-correlated single-photon counting (TCSPC) method on a home-built instrument based at RAL's CLF-Octopus facility. The HPTS was excited with 800 nm laser pulses (~100 ps), and two-photon absorption resulted in the emission from the photoexcited species around 510 nm.

**d. Data analysis:**

Data plotting and analysis were conducted in LABVIEW and MATLAB (R2021a).

## 2. Voigt profile analysis of transient absorption data:

To extract the time-dependent concentrations of the components of the photoprotolytic cycle of HPRS/EAF from the TR-IR data, we initially attempted single value decomposition deconvolution (SVD) to identify different components with distinguishable time-dependence. Significant overlap in the transient IR spectra between the different relevant species which co-exist together (ROH* and especially $RO^{*-}$ and $RO^{-}$) for a considerable period of time delay resulted in the SVD analysis giving components that did not separate into individual chemical species. The transient spectra at each delay time were instead fitted with model spectral profiles of ROH, $ROH^{*}$, $RO^{*-}$ and $RO^{-}$.

Voigt lineshapes were used for the spectral fitting.[12] Voigt lineshapes are a convolution of Gaussian and Lorentzian distributions. In the absence of detailed IR lineshape information for the ROH, $ROH^{*}$, $RO^{*-}$ and $RO^{-}$ species, Voigt lineshape fits encompass both forms of broadening.

The Voigt lineshapes are defined from a Lorentzian lineshape with a centre at $\hat{\nu}$ and width of $\gamma_L$ :

$$g_L(\nu-\hat{\nu},\gamma_L)=\frac{\gamma_L/\pi}{(\nu-\hat{\nu})^2+\gamma_L^2}$$

A Gaussian lineshape with centre at $\hat{\nu}$ and width of $\gamma_G$:

$$g_G(\nu-\hat{\nu},\gamma_G)=\frac{1}{\gamma_G}(\frac{ln2}{\pi})^{\frac{1}{2}}\exp[-ln2(\frac{\nu-\hat{\nu}}{\gamma_G})^2]$$

The Voigt lineshape is then calculated as:

$g_V(\nu - \hat{\nu}, \gamma_{G,}\gamma_L) = g_L \otimes g_G$, where $\otimes$ refers to convolution.

At early times (<1 ns), the dominating species in transient spectra are ROH* and $RO^{*-}$, whereas at very long delay times (>10 ns) the only species contributing is $RO^-$ therefore the transient spectra at extreme times were used as a guide to determine the Voigt profiles of these species.

### 3. Kinetic scheme analysis:

Scheme S1 describes the full photoprotolytic cycle of HPTS.

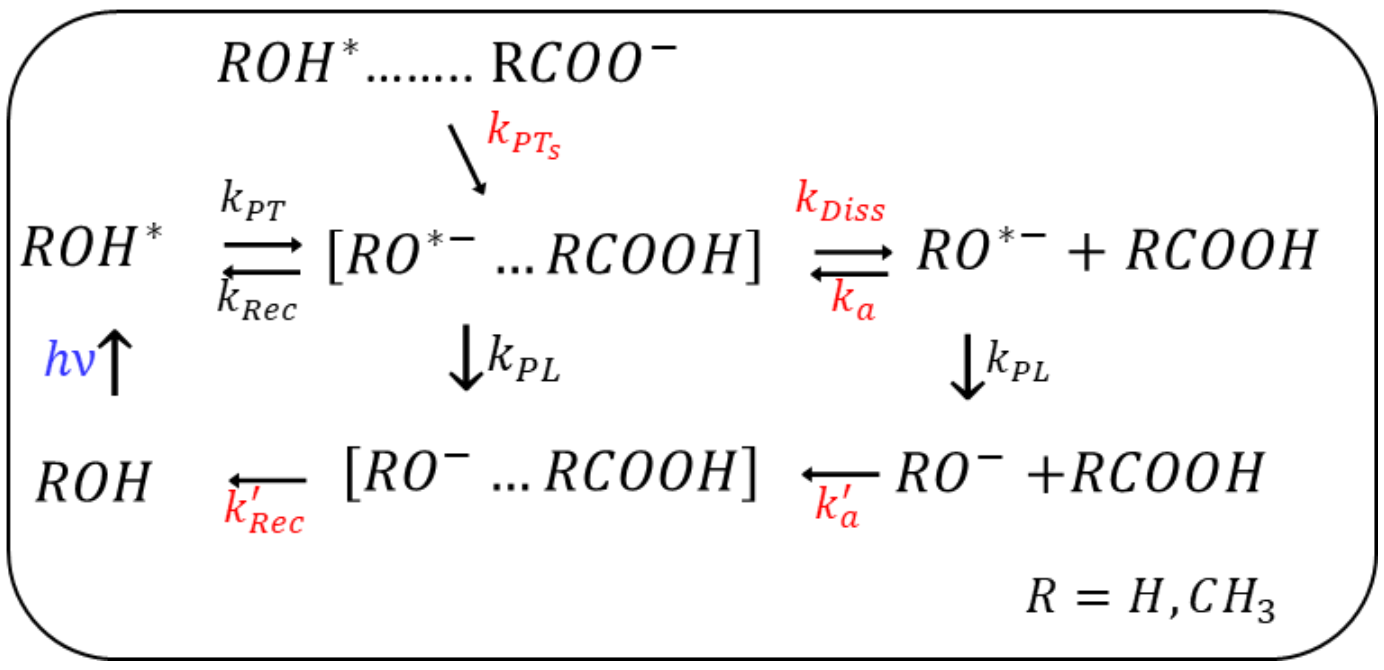

**Scheme S1.** A proposed photoprotolytic cycle of HPTS in water/acetate and EAF.

The rates of all the steps involved can be written as follows:

$$\frac{d[ROH^*]_1}{dt} = -k_{PT}\varphi_1[ROH^*] + k_{Rec}[RO^{*-} \dots RCOOH] \quad (1)$$

$$\frac{d[ROH^*]_2}{dt} = -k_{PT_s}\varphi_2[ROH^*] \quad (2)$$

$$\frac{d[RO^{*-} \dots RCOOH]}{dt} = (k_{PT}\varphi_1 + k_{PT_s}\varphi_2)[ROH^*] - (k_{diss} + k_{Rec} + k_{PL})[RO^{*-} \dots RCOOH] + k_a[RO^{*-}][RCOOH] \quad (3)$$

$$\frac{d[RO^{*-}]}{dt} = k_{Diss}[RO^{*-} \dots RCOOH] - k_a[RO^{*-}][RCOOH] - k_{PL}[RO^{*-}] \quad (4)$$

$$\frac{d[RO^-]}{dt} = k_{PL}[RO^{*-}] - k'_a[RO^-][RCOOH] \quad (5)$$

$$\frac{d[RCOOH]}{dt} = k_{Diss}[RO^{*-} \dots RCOOH] - k_a[RO^{*-}][RCOOH] - k'_a[RO^-][RCOOH] \quad (6)$$

$$\frac{d[RO^- \dots RCOOH]}{dt} = k'_a[RO^-][RCOOH] + k_{PL}[RO^{*-} \dots RCOOH] - k'_{Rec}[RO^- \dots RCOOH] \quad (7)$$

$$\frac{d[ROH]}{dt} = k'_{Rec}[RO^- \dots RCOOH] \quad (8)$$

$\varphi_1$ *and* $\varphi_2$ *refers to proportion of tight and loose complex, respectively.*

The coupled differential equations (1) – (8) were solved in Matlab (version R2021a) using solver "ode23s". The loosely bound proton population ($RO{-}{-}{-}H^{*-}$), $RO^{*-}$, $RO^-$ and formic acid populations were simultaneously fit through least-square analysis ("lsqcurvefit" in Matlab). We used "Multistart" (100 runs) to search for a global minimum during fitting. The standard errors were obtained through the bootstrapping method (resampling residuals for 100 runs). The overall $\chi^2$ values for HPTS/EAF system and DPTS/EAF (3D) system were 3.25 and 3.01, respectively.

## 4. Diffusion length analysis:

The diffusion length as described in the main text is estimated as follows:

The diffusion constant of Formic acid in water: $D_{HCOOH}$=1.5x10$^{-9}$ m$^2$/s

Diffusion coefficient of HPTS [13] (ROH) in water $D_{ROH}$=3.3x10$^{-10}$ m$^2$/s

The diffusion of the two species increases the total amount of diffusion:

$$D = D_{HCOOH} + D_{RO^-}$$

We assume $D_{RO^-} \sim D_{ROH}$

Given the net diffusion of the two species *D*, the average separation of the two species developing on time τ is:

$$L_D = \sqrt{D\tau}$$

The viscosity ($\eta$) for water: 1milliP.s    EAF: 23.1 milliP.s[14]

Using Stokes-Einstein relation $D \propto \frac{1}{\eta}$

Therefore, D in EAF=8x10$^{-11}$ m$^2$/s

$L_D$ in EAF:  $L_D$= 4.9 nm.

1 solvation shell is approx. 0.5 nm. [15]

Therefore the diffusion is about 8-10 solvation shells.